\documentclass[aps,icp,twocolumn,groupedaddress,longbibliography]{revtex4-1}
\usepackage[dvipsnames]{xcolor}
\usepackage{graphicx,amsmath,amssymb,color}
\begin{document}

\author{Joseph D. Dietz} 
\author{Robert S. Hoy}
\affiliation{Department of Physics, University of South Florida, Tampa, FL 33620 USA}
\email{rshoy@usf.edu}
\date{\today}

\title{Facile equilibration of well-entangled semiflexible bead-spring polymer melts}

\begin{abstract}
The widely used double-bridging hybrid (DBH) method for equilibrating simulated entangled polymer melts [R. Auhl \textit{et al.}, \textit{J.\ Chem.\ Phys.}\ \textbf{2003}, 119, 12718-12728] loses its effectiveness as chain stiffness increases into the semiflexible regime because the energy barriers associated with double-bridging Monte Carlo moves become prohibitively high.
Here we overcome this issue by combining DBH with the use of core-softened pair potentials.
This reduces the energy barriers substantially, allowing us to equilibrate melts with $N \simeq 40 N_e$ and chain stiffnesses all the way up to the isotropic-nematic transition using simulations of no more than 100 million timesteps. 
For semiflexible chains, our method is several times faster than standard DBH; we exploit this speedup to develop improved expressions for Kremer-Grest melts' chain-stiffness-dependent Kuhn length $\ell_K$ and entanglement length $N_e$.
\end{abstract}
\maketitle

\section{Introduction}
\label{sec:intro}

Equilibration of simulated entangled polymer melts is a longstanding challenge.
The longest relaxation time for a single $N$-mer chain, i.e.\ the disentanglement time $\tau_{\rm d}$ after which the chain has escaped from its original tube, is $\tau_{\rm d} = \tau_0 (N/N_e)^{3.4}$, where $N_e$ is the melt's entanglement length and $\tau_0$ is a microscopic time scale.\cite{doi88}
Since chains' mean radius of gyration $R_{\rm g}$ and end-end distance $R_{\rm ee}$ scale as $N^{1/2}$, and interactions of chains with their own periodic images produce spurious results, the minimum simulation cell side length $L_{\rm min}$ also scales as $N^{1/2}$, and thus the minimum simulation volume $V_{\rm min}$ scales as $N^{3/2}$.
For a melt of fixed monomer number density $\rho$, this means that the minimum number of monomers $N_{\rm mm} \sim N^{3/2}$.
Thus the total computational effort required to equilibrate a polymer melt is\cite{dunweg98}
\begin{equation}
E \sim V_{\rm min} \tau_{\rm d} \sim N^{3/2} \left( \displaystyle\frac{N}{N_e} \right)^{3.4} \tau_0 \sim N^{4.9} \tau_0,
\end{equation}
\textit{provided the polymer chains follow physically realistic dynamics} (i.e.\ reptation).
This nearly-$N^5$ scaling limited early molecular dynamics (MD) and Monte Carlo (MC) simulations of polymer melts to at-most-weakly-entangled chains with $N$ no more than a few $N_e$,\cite{kremer90,paul91} and still limits simulations of well-entangled chains to durations of no more than a few $\tau_d$ even on modern supercomputers.\cite{hou10,oconnor19}

Many problems of current interest can be addressed through simulations of duration much shorter than $\tau_{\rm d}$.
However, this still leaves the problem of equilibration; simulations must begin with a reasonable-guess initial configuration and then run for a time sufficient to ensure it reaches equilibrium before the ``production'' run can begin.
If the chains follow physically realistic dynamics, this requires an equilibration run of duration $\gtrsim \tau_d$.
Fortunately, these runs need not follow physically realistic dynamics, and several modern equilibration algorithms exploit this fact.\cite{generationMK,karayiannis02,karayiannis02b,auhl03,bisbee11,sliozberg12,bobbili20,zhang14,zhang15,moreira15,svaneborg16,lemarchand19,generationMKupdate,zhang19,hsu20,tubiana21}
The simpler algorithms fall into two basic categories: core-softening\cite{sliozberg12,bobbili20} and topology-changing.\cite{karayiannis02,karayiannis02b,auhl03,bisbee11}
Both approaches greatly speed up diffusive equilibration by eliminating the constraints on chains' transverse motion (and hence their slow reptation dynamics), but both have inherent limitations.

Core-softening speeds up diffusive equilibration by allowing chains to cross.
However, it also produces a local melt structure that can differ substantially from the equilibrium structure for the final interactions.
For example, the cluster-level structure of hard-sphere liquids is more ordered than that of WCA  liquids at the same temperature and density because the latter have more free volume.\cite{taffs10,royall15} 
The degree to which comparable many-body effects occur in polymer melts has not been explored, but it is reasonable to assume that intermediate-scale melt structure couples rather strongly to monomer-scale structure via chains' connectivity and semiflexibility.
Any resulting structural differences must be annealed out via diffusion after the potentials are restored to their final form, and the minimum required annealing time is often \textit{a priori} unclear, particularly for the higher-order structural correlations that determine $N_e$.\cite{graessley81,kavassallis87,morse98,milner20,hoy20}

Topology-changing methods speed up diffusive equilibration by eliminating permanent chain connectivity.
These methods maintain monomer-scale structure while bringing larger-scale intrachain and interchain structure to equilibrium, but perform poorly when  covalent-bond and/or bond-angular interactions are stiff.
For example, the well-known double-bridging hybrid (DBH) algorithm\cite{auhl03} reduces the longest single-chain relaxation time to $\tau_{\rm max} \sim (N/N_e) \tau_0$, and thus the total computational effort to $E \sim N^{5/2} \tau_0$, by performing periodic  bond/angle-swapping MC moves (Figure \ref{fig:schem}) during a MD simulation.
This method works well for flexible and nearly-flexible chains, even very long ones.\cite{hou10,svaneborg16} 
For semiflexible chains, however, the high energy barriers associated with these MC moves make the prefactors of the $\tau_{\rm max} \sim (N/N_e)\tau_0$ scaling prohibitively large.

Here we overcome this $\tau_{\rm max}$-prefactor issue by introducing a novel equilibration algorithm that combines DBH with core-softening.
By beginning with soft pair and bond interactions, and slowly stiffening them until they reach their final functional forms while keeping the equilibrium bond length constant, we are able to equilibrate large ($4\times10^5$-monomer) systems with $N \simeq 40 N_e$ and chain stiffnesses all the way up to the isotropic-nematic transition, using simulations that last no more than $\mathcal{N}_t^{\rm max} = 10^8$ MD timesteps.
The required equilibration times are several times shorter than for standard DBH.
We exploit this speedup to develop improved expressions for Kremer-Grest melts' chain-stiffness-dependent $N_e$ and Kuhn length $\ell_K$.

\section{Equilibration Method}
\label{sec:methods}

\subsection{Interaction potentials}
\label{subsec:potentials}

We demonstrate our method using the semiflexible version of the widely used Kremer-Grest (KG) bead-spring polymer model.\cite{kremer90}
The standard nonbonded-pair, covalent-bond, and bond-angular interactions for this model are respectively the truncated and shifted Lennard-Jones potential
\begin{equation}
U_{\rm LJ}(r) = 4\epsilon\left[ \left(\displaystyle\frac{\sigma}{r} \right)^{12} -  \left(\displaystyle\frac{\sigma}{r} \right)^{6} - \left(\displaystyle\frac{\sigma}{r_c} \right)^{12} +  \left(\displaystyle\frac{\sigma}{r_c} \right)^{6} \right], 
\label{eq:stdLJ}
\end{equation}
where $\epsilon$ and $\sigma$ are characteristic energy and length scales, $r$ is the intermonomer distance, and $r_c = 2^{1/6}\sigma$ is the cutoff radius, the FENE potential\cite{bird77}
\begin{equation}
U_{\rm FENE}(r) = U_{\rm LJ}(r) -\displaystyle\frac{ k R_0^2}{2} \ln\left[1- \left( \displaystyle\frac{r}{R_0} \right)^2\right]
\label{eq:stdFENE}
\end{equation}
where $k = 30\epsilon\sigma^{-2}$ and $R_0 = 1.5\sigma$, and   
\begin{equation}
U_{\rm ang}(\theta) = \kappa [1 - \cos(\theta)],
\label{eq:stdang}
\end{equation}
where $\theta = \cos^{-1}(\hat{b}_i\cdot \hat{b}_{i+1})$ is the angle between consecutive bond vectors $\vec{b}_i$ and $\vec{b}_{i+1}$.\cite{faller99}
This model was recently shown by Svaneborg, Everaers and colleagues to accurately capture the dynamics of a wide variety of commodity polymer melts when $\epsilon$, $\sigma$, and $\kappa$ are taken as adjustable parameters.\cite{everaers20,svaneborg20}

\begin{figure}[h]
\includegraphics[width=2.875in]{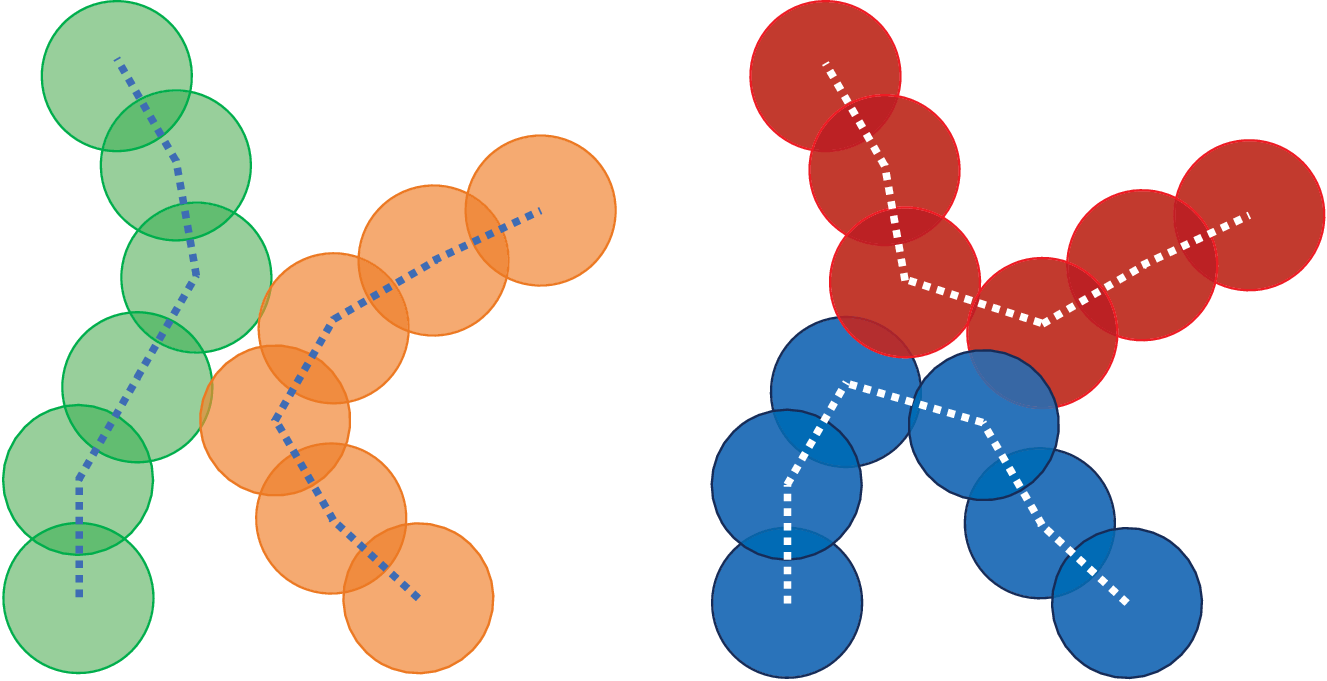}
\caption{A DBH swap move converts a low-energy configuration (green and orange chains; $E = E_0$) into a higher-energy configuration (red and blue chains; $E = E_0 + \Delta U_{\rm bond} + \Delta U_{\rm ang}$) by replacing two covalent bonds and four angles.\cite{karayiannis02,karayiannis02b,auhl03}  This move is accepted with probability $\exp(-[\Delta U_{\rm bond} + \Delta U_{\rm ang}]/k_B T)$, while the reverse move is always accepted (assuming $\Delta U = \Delta U_{\rm bond} + \Delta U_{\rm ang} > 0$).  When the standard KG model is employed in its standard temperature range ($k_B T \leq \epsilon$), the forward moves become prohibitively unlikely for $\kappa \gtrsim 2.5\epsilon$.}
\label{fig:schem}
\end{figure}

Equilibration of entangled KG melts becomes increasingly computationally expensive as chain stiffness increases.
For the standard temperature ($T = \epsilon/k_B$), double-bridging algorithms\cite{karayiannis02,karayiannis02b,auhl03} are effective only up to $\kappa \simeq 2.5\epsilon$ because for larger $\kappa$ the larger energy barriers for angle-swapping makes their MC acceptance rates prohibitively low; see Figure \ref{fig:schem}.
To reduce these energy barriers, we initially employ a core-softened pair potential that allows thermally activated chain-crossing but preserves large-scale chain structure, and then gradually harden it until the final potential employed in the production runs [i.e.\ the standard Lennard-Jones potential $U_\textrm{LJ}(r)$] is operative.

A natural choice for a core-softened pair potential is the generalized Lennard-Jones (Mie) potential\cite{mie03} 
\begin{equation}
U_\textrm{LJ}^{n-m}(r) = \displaystyle\frac{\epsilon}{n-m} \left[ 2^{n/6} m \left(  \displaystyle\frac{\sigma}{r} \right)^n - 2^{m/6} n \left(  \displaystyle\frac{\sigma}{r} \right)^m \right],
\label{eq:genLJ}
\end{equation}
which is symmetric with respect to an exchange of $n$ and $m$.
The standard LJ potential is recovered for $n=2m=12$.
The factors of $2^{n/6}$ and $2^{m/6}$ place the minimum of $U_\textrm{LJ}^{m-n}$ at $r = 2^{1/6}\sigma$ (the standard Lennard-Jones value) for any $n$ and $m$.
However, $U_\textrm{LJ}^{n-m}(r)$ is not optimal to use during equilibration of polymers with attractive pair interactions ($r_c > 2^{1/6}\sigma$) because there is no advantage to be gained by modifying the attractive ($r > 2^{1/6}\sigma$) portion of the pair potential.
Therefore we proposed\cite{hoy20} the modified generalized-LJ potential
\begin{equation}
U_\textrm{MLJ}^{n}(r) = \left\{ \begin{array}{l@{\quad}l}
U_\textrm{LJ}^{n-6}(r), &  r \leq 2^{1/6}\sigma\\
\\
U_\textrm{LJ}^{12-6}(r), & r \geq 2^{1/6}\sigma
\end{array}\right .
\label{eq:upair}
\end{equation}
for use in equilibrating semiflexible bead-spring polymer melts.
We employ a truncated-and-shifted version of $U_\textrm{MLJ}^{n}(r)$  in the equilibration runs described below.

\begin{figure}[h]
\includegraphics[width=3in]{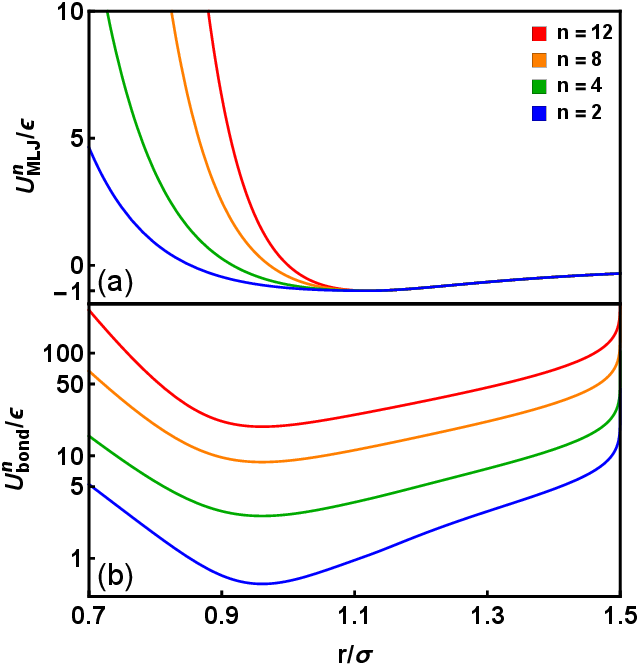}
\caption{Potentials used during equilibration.  Panel (a):\ the pair potential $U_\textrm{MLJ}^{n}(r)$ (Eq.\ \ref{eq:upair}).  Panel (b):\ the bond potential $U_\textrm{bond}^{n}(r)$ (Eq.\ \ref{eq:ubondmn}).  While this figure shows $U_\textrm{MLJ}^{n}(r)$ and $U_\textrm{bond}^{n}(r)$ for $r_c = 2^{7/6}\sigma$ to illustrate the attractive part of the potentials, the simulations described below use $r_c = 2^{1/6}\sigma$.  Note that the vertical axis of panel (b) is log-scale.}
\label{fig:pots}
\end{figure}

Our method works by beginning with soft repulsive pair interactions ($n \ll 12$), adjusting $k$ to keep the mechanical-equilibrium bond length $\ell_0$ at its standard value,\cite{mechvstherm} and then performing double bridging moves while gradually increasing $n$.
In the standard KG model, $k = 30\epsilon/\sigma^2$ is the solution to
\begin{equation}
\displaystyle\frac{\partial}{\partial r} U_\textrm{FENE}(r) \big{ |}_{r =\ell_0} = 0 ,
\label{eq:stdl0}
\end{equation}
where $\ell_0 \equiv 0.960897\sigma$.
When the pair potential is $U_\textrm{MLJ}^{n}$, the FENE spring constant producing an mechanical-equilibrium bond length equal to $\ell_0$ is instead given by the solution to
\begin{equation}
\displaystyle\frac{\partial}{\partial r} \left( U_\textrm{MLJ}^{n}(r)  - \displaystyle\frac{  k_n  R_0^2}{2} \ln\left[1- \left( \displaystyle\frac{r}{R_0} \right)^2\right]  \right)_{r=\ell_0} = 0 ,
\label{eq:modl0}
\end{equation}
i.e.
\small
\begin{equation}
k_n =  \displaystyle\frac{6 n \left(R_0^2 - \ell_0^2 \right)  \left( 2^{n/6} \ell_0^{-n} - 2 \ell_0^{-6}   \right)}{ (n-6)R_0^2 \ell_0^2} \displaystyle\frac{\epsilon}{\sigma^2} . 
\label{eq:kfenemn}
\end{equation}
\normalsize
Then the overall interaction potential between covalently bonded monomers is
\begin{equation}
U_\textrm{bond}^{n}(r) = U_\textrm{MLJ}^{n}(r) - \frac{ k_n R_0^2}{2}  \ln\left[1- \left(\frac{r}{R_0}\right)^2\right].
\label{eq:ubondmn}
\end{equation}

As shown in Figure \ref{fig:pots}, $U_\textrm{MLJ}^{n}$ and $U_\textrm{bond}^{n}$ both soften substantially when $n$ is decreased with $m$ held fixed.
Below, we will show that beginning runs with $n = 2$ and gradually incrementing $n$ through the set 
\begin{equation}
\Xi = \{2,3,4,5,5.75,6.5,7,8,9,10,11,12\}
\label{eq:nset}
\end{equation}
 produces orders-of-magnitude increases in the average DBH swap success rate, which in turn leads to much faster equilibration.
Note that this protocol does \textit{not} reduce the angular component of the energy barrier  ($\Delta U_{\rm ang}$) for a given double-bridging move (Fig.\ \ref{fig:schem}).  
Instead, the softer bonded interactions (Eq.\ \ref{eq:kfenemn}-\ref{eq:ubondmn}) substantially reduce ($\Delta U_{\rm bond}$), while the softer pair interactions (Eqs.\ \ref{eq:genLJ}-\ref{eq:upair})  allow chain crossing (for $n \lesssim 8$) as well as deeper interpenetration of  swap-candidate pairs that in turn reduce $\langle \Delta U_{\rm ang} \rangle$.

\subsection{Detailed protocol}
\label{subsec:details}

Any polymer-melt-equilibration protocol begins by generating an initial state; better initial states allow for shorter equilibration runs.
Svaneborg \textit{et al.}\ recently showed\cite{svaneborg20} that the equilibrium Kuhn length of $0 < \kappa \lesssim 2.5\epsilon$ KG chains at the standard melt temperature ($T = \epsilon/k_B$) is well-approximated by\cite{eq12}
\begin{equation}
\begin{array}{lcl}
\displaystyle\frac{\ell_K}{\ell_0} & = & \displaystyle\frac{2 \kappa/\epsilon + \exp(-2 \kappa/\epsilon)  -1 }{1 - \exp(-2 \kappa/\epsilon) (2 \kappa/\epsilon + 1)} \\
\\
& & +\ 0.77\left[ \tanh( - 0.03\kappa^2/\epsilon^2 - 0.41\kappa/\epsilon  + 0.16) + 1 \right].
\end{array}
\label{eq:svanekuhn}
\end{equation}
The first term in Eq.\ \ref{eq:svanekuhn} is the standard Flory term,\cite{flory53} and the second term is an empirical correction term associated with the long-range bond-orientational correlations found in dense polymer melts.\cite{wittmer07}

For any given $\kappa$, we generate initial states by placing $N_{\rm ch}$ freely-rotating (FR) $N$-mer chains with bond angle 
\begin{equation}
\theta = \theta^*(\kappa) = \cos^{-1}\left[  \displaystyle\frac{\ell_K(\kappa)/\ell_0 + 1}{\ell_K(\kappa)/\ell_0 - 1 } \right]
\label{eq:thetainit}
\end{equation}
in a cubic cell of volume $V = N_{\rm ch}N/\rho$, where $\ell_K(\kappa)$ is given by Eq.\  \ref{eq:svanekuhn}.
FR chains with $\theta = \theta^*(\kappa)$ are guaranteed to have the correct $\langle \ell_K(\kappa) \rangle$ in the large-$N_{\rm ch}$, large-$N$ limit.
To minimize finite-sampling errors, we fit the average chain statistics $\langle R^2(n)\rangle$ the WLC formula
\begin{equation}
\displaystyle\frac{\langle R^2(n)\rangle}{n\ell_0} = \ell_K \left\{1- \displaystyle\frac{\ell_K}{2n\ell_0} \left[ 1-\exp\left(-\displaystyle\frac{2n\ell_0}{\ell_K}\right)\right]\right\},
\label{eq:lk}
\end{equation}
at intermediate $n$ (e.g.\ $N/4 \leq n \leq N/2$) and reject the state if the fit $\ell_K$ deviates from the prediction of Eq.\  \ref{eq:svanekuhn} by more than 2\%.
All results presented in Sections \ref{subsec:details}-\ref{sec:perf} are for $N_{ch} = 1000$ and $N = 400$.

Once we have generated a satisfactory initial state, we set $n = 2$ in Eqs.\ \ref{eq:upair} and \ref{eq:ubondmn} and begin the MD simulation.
All monomers have mass $\tilde{m}$, and all MD simulations are performed using LAMMPS.\cite{plimpton95}
Because $n = 2$ FENE bonds are much weaker than their $n = 12$ counterparts (Fig.\ \ref{fig:pots}), a very small initial timestep $dt_{\rm init}$ is required to avoid overstreched-bond crashes arising from initial monomer overlap.
We choose $dt_{\rm init} \sim .001\tau_{\rm LJ}$, where $\tau_{\rm LJ} = \sqrt{\tilde{m} \sigma^2/\epsilon}$ is the standard Lennard-Jones time unit.
After ramping the MD timestep up to its final value ($dt = \tau/125$) over a few hundred $\tau_{\rm LJ}$, we activate the MC double-bridging moves.

In our simulations, one double-bridging move (Fig.\ \ref{fig:schem}) is attempted per every $10^3$ monomers per $\tau_{\rm LJ}$, i.e.\ moves are attempted at a rate $R_{\rm MC} = 10^{-3}N_{\rm ch}N/\tau_{\rm LJ}$.
Moves are attempted only for pairs of chains that:\ (i) allow for a bond-swap that preserves the length of both chains (i.e., keeps the melt monodisperse\cite{auhl03}), and (ii) do not involve creation of a new covalent bond with length $\ell > 1.3\sigma$.

\begin{figure}[htbp]
\includegraphics[width=3in]{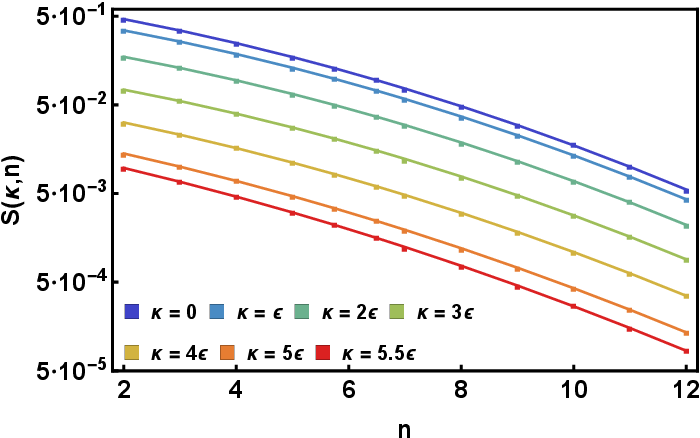}
\caption{Bond-swap success rates (symbols) and fits to $S(\kappa,n) = a'\exp[ -b' \kappa/\epsilon -  d'n -e'n^2]$ (curves), with $a' = 1.24  \pm 0.036$ , $b' = 0.77 \pm 0.0096$, $d' =  0.24 \pm 0.038$, and  $e' = 0.015 \pm 0.0016$,  for selected $\kappa$. }
\label{fig:swap1}
\end{figure}

Before describing our protocol any further, it is worthwhile to examine how the MC double-bridging success rate varies with $\kappa$ and $n$.
Figure \ref{fig:swap1} shows success rates for $2 \leq n \leq 12$ and $0 \leq \kappa \leq 5.5\epsilon$.
We find that the success rates for $\kappa \gtrsim \epsilon$ and $n \lesssim 9$ are approximately fit by 
\begin{equation}
S(\kappa,n) = a\exp\left[ -b \displaystyle\frac{\kappa}{\epsilon} -  dn \right] ,
\label{eq:srate}
\end{equation}
where $a \simeq 1.6 \pm 0.037$, $b \simeq 0.77 \pm 0.015$,  and $d \simeq 0.35 \pm .01$.
The large value of $b$ indicates why standard [$n=12$-only] DBH  equilibration\cite{auhl03} becomes impractical for  $\kappa > 2.5\epsilon$.
On the other hand, the large value of $d$ indicates that combining DBH with core-softening is an effective way of addressing this issue.

Auhl \textit{et al} suggested that a fixed number of successful MC swaps per monomer is necessary to equilibrate a polymer melts.\cite{auhl03}.
Because DBH keeps systems strictly monodisperse,\cite{karayiannis02b} the probability that a swap candidate lies in the vicinity of a given bond scales as $1/N$, which suggests that the minimum total duration of the equilibration run ($\tau_{\rm eq})$ scales at least linearly with $N$ for $N \gg N_e$.
Combined with the results shown in Fig.\ \ref{fig:swap1}, this suggests that $\tau_{\rm eq}$ should be at least 
\begin{equation}
\tau_{\rm fix}(\kappa) = \displaystyle\frac{fN}{S(\kappa,0)} \tau_{\rm LJ} \equiv \displaystyle\frac{fN}{a}  \exp\left[ b \displaystyle\frac{\kappa}{\epsilon} \right] \tau_{\rm LJ},
\label{eq:taufix}
\end{equation}
where $f$ is a numerical prefactor.
Since we wish to develop an efficient equilibration algorithm, a reasonable criterion for $f$ is that it must allow well-entangled 
($N  \gg  N_e$) systems to be equilibrated using runs of $\mathcal{N}_t \leq \mathcal{N}_t^{\rm max} = 100$ million timesteps.
For $N = 400$, meeting this criterion for systems near the onset of local nematic order ($\kappa = 5.5\epsilon$\cite{faller99}) requires $f \leq f_0 = 46.3354$.
We will show below that $f = f_0$ is indeed sufficient to equilibrate these $\kappa = 5.5\epsilon$ melts in $\mathcal{N}_t^{\rm max}$-timestep runs, while smaller-$\kappa$ melts can be equilibrated much faster.

To ensure equilibration of interchain structure on all length scales, $\tau_{\rm eq}$ should also be at least $2\tau_e$, where $\tau_e = \tau_R(N_e)$ is the Rouse time of a typical entangled segment.
At least half (i.e.\ $\tau_e$) of this time should employ the final ($n = 12$) pair interactions, to insure  that the local inter- and intra-chain structure of the melt reflects these interactions.
Svaneborg and Everaers recently showed\cite{svaneborg20} that 
\begin{equation}
\tau_e(\kappa)  = N_{\rm eK}^2(\kappa) \tau_k(\kappa) =  \displaystyle\frac{12 \eta_K }{\pi ^2 k_B T}  N_{e}^2 (\kappa)\ell_K (\kappa),
\label{eq:taue1}
\end{equation}
where $N_{\rm eK} \equiv N_e \ell_0/\ell_K$ is the number of Kuhn segments per entanglement, the  Kuhn time $\tau_K$ is the time over which Kuhn segments diffuse a distance comparable to their own size, and $\eta_K$ is the local, Kuhn-segment-scale melt viscosity.
We ignored the $\mathcal{O}(1)$ prefactor in the rightmost term of Eq.\ \ref{eq:taue1} and employed the approximation $\tau_{e}(\kappa) =  N_{e}^2 (\kappa)[ \ell_K (\kappa)/\ell_0]\tau_{\rm LJ}$, with values of $\ell_K(\kappa)$ calculated from Eq.\ \ref{eq:svanekuhn} and an initial estimate $N_e(\kappa) = 77\exp(-\kappa/1.1) + 9.4$.
This allowed us to estimate $\tau_e(\kappa)$ and thus $\tau_{\rm eq}(\kappa,N)$; a more accurate expression for $N_e(\kappa)$ is given in Section \ref{sec:equilstruct}.

\begin{figure}[h]
\includegraphics[width=3in]{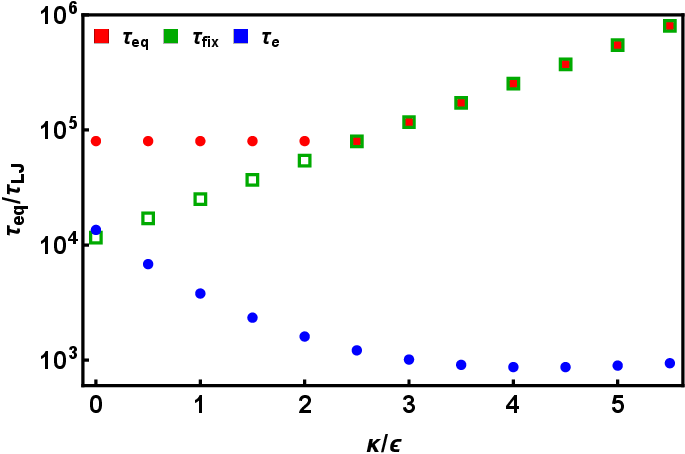}
\caption{Total equilibration runtime $\tau_{\rm eq}(\kappa)$ tor $N = 400$ Kremer-Grest melts.  For $dt=.008\tau$, these runs require 10 million timesteps for $\kappa \leq 2.5\epsilon$ and 100 million for systems near the onset of local nematic order ($\kappa = 5.5\epsilon$).  Values of $\tau_e(\kappa)$ and $\tau_{\rm fix}(\kappa)$ are also shown to provide context.}
\label{fig:tauswap}
\end{figure}

Combining the above arguments with an additional empirical criterion that equilibration runs' duration should be at least $200N\tau$ suggests that a reasonable duration for the entire run is
\begin{equation}
\tau_{\rm eq}(\kappa) = \max[2\tau_e(\kappa) ,\tau_{\rm fix}(\kappa),200N\tau].
\label{eq:maxdur}
\end{equation}
Values of $\tau_{\rm fix}$, $\tau_e$ and $\tau_{eq}$ obtained from Eqs.\ \ref{eq:taufix},  \ref{eq:taue1}, and \ref{eq:maxdur} are given in Figure \ref{fig:tauswap}.
For context, note that the estimated values of $\tau_d = 3(N/N_e)^{3.4}$ for these systems (using the $N_e$ values given in Section \ref{sec:equilstruct}) range from $\sim 1.4\cdot 10^7 \tau$ for $\kappa = 0$ to $\sim 7.1\cdot 10^8 \tau$ for $\kappa = 5.5\epsilon$.

The next question to answer is:\ how should the run be divided amongst the various values of $n$ (Eq.\ \ref{eq:nset})?
Clearly, spending more time at smaller $n$ allows more swaps to be executed, which improves equilibration.
However, as discussed above, the $n = 12$ portion of the runs should last at least $\tau_e$.
After some trial and error, we devised two \textit{ad hoc} division schemes that satisfy both criteria:\ one for $\kappa \leq 2.5\epsilon$ and another for $\kappa > 2.5\epsilon$.

For $\kappa < 2.5\epsilon$, total run duration is $200N\tau$.
We assign the last $2\tau_e$ of the run to $n = 12$, and the remaining $\tau^*(\kappa) = 200N\tau - 2\tau_e(\kappa)$ to $n < 12$.
The time assigned to each $n$ is given by
\begin{equation} 
\tau_i(\kappa) = \displaystyle\frac{\Xi_i^{-1/2}}{\displaystyle\sum_{j = 1}^{11} \Xi_j^{-1/2} } \tau^*(\kappa)
\label{eq:scheme1portions}
\end{equation}
for $i = 1,2,...,11$, where $\Xi$ was defined in Eq.\ \ref{eq:nset}.

For $\kappa \geq 2.5\epsilon$, the total run duration is $\tau_{\rm fix}(\kappa)$.
The time assigned to each $n$ is given by
\begin{equation} 
\tau_i(\kappa) = \displaystyle\frac{\Xi_i^{-1/2}}{\displaystyle\sum_{j = 1}^{12} \Xi_j^{-1/2} }  \tau_{\rm fix}(\kappa)
\label{eq:scheme2portions}
\end{equation}
for $i = 1,2,...,12$.
For the $N = 400$ systems considered here (as well as all larger $N$), all systems have $\tau_{12} \geq \tau_e$, so our $n=12$ requirement is automatically satisfied.

In the following section, we describe the performance of this algorithm in detail.
All results are averaged over 10 independently prepared systems.

\section{Performance of the Algorithm}
\label{sec:perf}

\begin{figure}[h]
\includegraphics[width=3in]{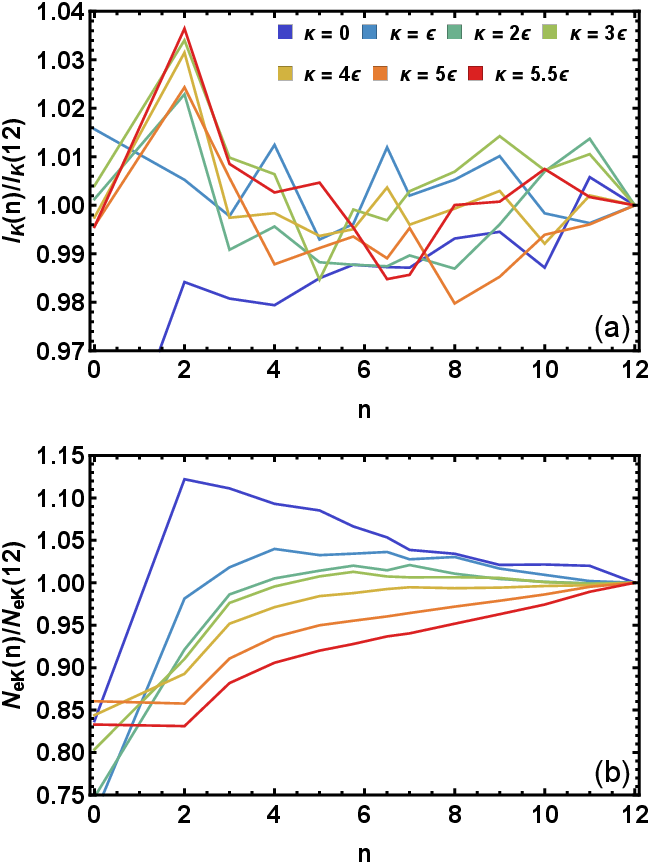}
\caption{Kuhn length $\ell_K$ [panel (a)] and number of Kuhn segments per entanglement $N_{\rm eK}$ [panel (b)] at the end of each $n$-step, divided by their values at the end of the equilibration run (i.e.\ at the end of its $n = 12$ portion).  The $``n = 0$'' data shows values of these quantities at the end of the timestep-ramp (Sec.\ \ref{subsec:details}), i.e.\ their values immediately before bond-swapping is turned on.}
\label{fig:convwithn}
\end{figure}

Figure \ref{fig:convwithn} shows how large-scale intrachain structure and entanglement vary with $n$ over the course of equilibration runs.
Panel (a) shows data for selected systems' Kuhn lengths $\ell_K$, obtained by fitting their large-$n$ chain statistics to Eq.\ \ref{eq:lk}.
Here $\ell_K(n)$ is the Kuhn length at the end of each $n$-step, while $\ell_K(12)$ is the Kuhn length at the end of the equilibration run.
$\ell_K$ decreases slightly with increasing $n$ for most systems, primarily because the softer bond potentials at smaller $n$ (Eq.\ \ref{eq:kfenemn}) lead to a slightly larger thermodynamic-equilibrium bond length $\langle \ell_0 (T) \rangle$, i.e.\ our protocol produces a small negative $d\langle \ell_0 (T) \rangle/dn$.
However, because we optimized our initial intrachain structure using Eqs.\ \ref{eq:svanekuhn}-\ref{eq:thetainit}, the fractional changes in $\ell_K$ during the runs are no more than $\sim 3\%$ for $\kappa > 0$.
The larger initial deviation for $\kappa = 0$ is present because the $\kappa \to 0$ limit of Eq.\ \ref{eq:svanekuhn}, which we used to generate our initial states, does not correctly predict these systems' equilibrium $\ell_K$; it can be eliminated by generating initial states with the correct $\ell_K \simeq 1.82\sigma$.

Entanglement lengths measured by topological analyses are a highly sensitive method of equilibration because they depend sensitively on both intrachain and interchain structure.\cite{hoy05,moreira20}
We calculate systems' $N_e$ using primitive path analysis\cite{everaers04} and the formula recently proposed by Svaneborg and Everaers:\cite{svaneborg20}
\begin{equation}
\displaystyle\frac{2N_{\rm eK} + \exp(-2N_{\rm eK})  - 1}{2N_{\rm eK}^2} =  \left( \displaystyle\frac{L_{pp}}{L} \right)^2 .
\label{eq:carNe}
\end{equation}
Here $L_{pp}$ is the average length of the primitive paths obtained by PPA, $L = (N-1)\ell_0$ is chains' normal contour length, and $N_{\rm eK}$ is the number of \textit{Kuhn segments} per entanglement.
Then $N_e$ is simply $N_{\rm eK} \ell_K/\ell_0$.
Unlike previous $N_e$-estimators,\cite{everaers04,kroger05,hoy09} Eq.\ \ref{eq:carNe} properly accounts for chain semiflexibility, i.e.\ for chains and primitive paths that are not random-walk-like between entanglements.

Panel (b) shows how these systems' reduced entanglement lengths $N_{\rm eK}$ evolve during the equilibration runs.
The initial states are far more entangled than the final states, as is typical for systems created by randomly placing and orienting chains within the simulation cell.\cite{hoy05}
The $n = 2$ stages of the runs, which yield the most successful swaps and the most chain crossing of any stage (Figs.\ \ref{fig:pots}-\ref{fig:swap1} and Eqs.\ \ref{eq:scheme1portions}-\ref{eq:scheme2portions}), produce very substantial disentanglement for most $\kappa$.
As $n$ continues to increase, $N_{\rm eK}$ values  progress steadily towards their final values, at rates that decrease with increasing $n$, for all systems.
This trend suggests that our use of core-softened potentials does not significantly perturb systems away from their equilibrium structure, and is consistent with previous studies that have employed core-softening to promote chain-crossing.\cite{sliozberg12,bobbili20}
To verify this hypothesis, and as a further check for equilibration on all length scales, we calculated the structure factor $S(q)$ at the end of each $n$-step, and found that it evolves monotonically towards equilibrium as $n$ increases, for all $q$ and all $\kappa$.
Further details are given in Appendix A.

\begin{figure}[h]
\includegraphics[width=3in]{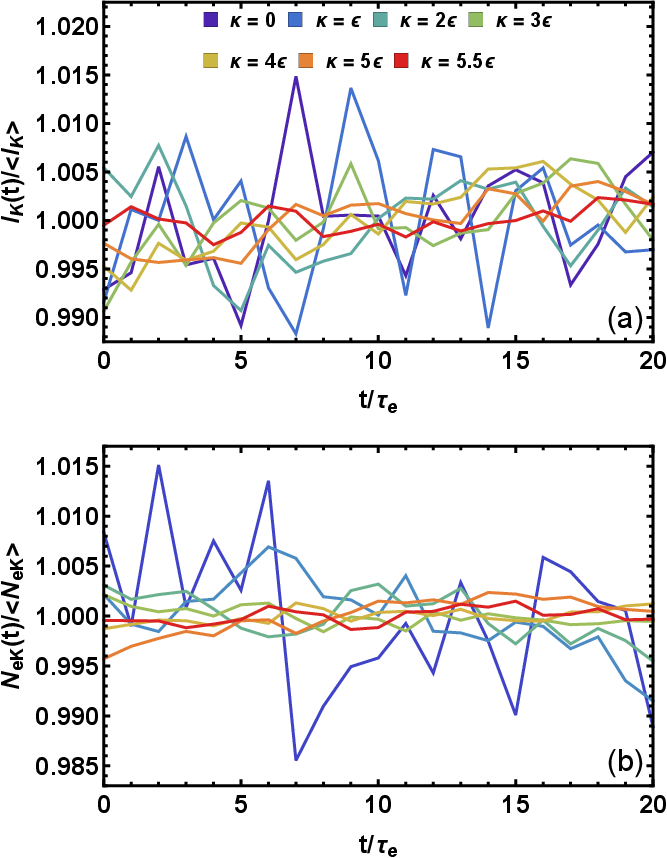}
\caption{Temporal fluctuations in $\ell_K$ and $N_{\rm eK}$ during the extended $n = 12$ runs.}  
\label{fig:extn12}
\end{figure}

Next we check whether all systems have in fact reached equilibrium by the end of these runs.
We performed long follow-up $n=12$ runs and measured the fractional changes in systems' $\ell_K$ and $N_{\rm eK}$ values---with respect to their time-averaged mean---as a function of time $t$.  
Bond-swapping was continued during these $20 \tau_e$-long runs to produce greater decorrelation of systems' structure.
Figure \ref{fig:extn12} shows ensemble-averaged results for $\ell_K(t)/\langle \ell_K \rangle$ and $N_{\rm eK}(t)/\langle N_{\rm eK} \rangle$.
Temporal fluctuations in both quantities are centered about 1, with magnitudes below $0.5\%$ for $\kappa \geq \epsilon$.
The flexible ($\kappa = 0$) melts have somewhat larger fluctuations because they have only $\sim 5$ entanglements per chain, but also show no systematic trends with time.
For all $\kappa$, $|\ell_K(20\tau_e) - \ell_K(0)|$ and $|N_{\rm eK}(20\tau_e)- N_{\rm eK}(0)|$ are small compared to  \textit{single-system}, non-ensemble-averaged fluctuations in these quantities.
We therefore conclude that all systems are (at least) asymptotically close to equilibrium at the end of their equilibration runs.

The protocol described above is more complicated than standard DBH equilibration; supporting its use requires demonstrating that it produces a significant speedup.
In Table \ref{tab:swapspeed} we show that this is indeed the case.
We list the number of successful bond swaps per monomer over the course of the equilibration runs for each $\kappa$, and estimated speedups for our protocol relative to DBH runs that employ only the standard ($n = 12$) interactions.
The speedup ratios $R$ can be predicted from the details of the algorithm (as described above) using the formula
\begin{equation}
R(\kappa) = \displaystyle\frac{ \displaystyle\sum_{i = 1}^{i_{\rm max}} \tau_i S(\kappa,\mathcal{N}_i) }{ \tau_{\rm fix}(\kappa) S(\kappa,12) }.
\end{equation}
Values of $R$ increase monotonically with $\kappa$, and are above $20$ for semiflexible ($\kappa \gtrsim 2\epsilon$) chains.

\begin{table}[h]
\begin{ruledtabular}
\caption{Successful DBH swaps per monomer and estimated speedup ratios provided by our method.  The speedup $R$ is equal to  $\mathcal{N}_t^{\rm n=12}/\mathcal{N}_t$, where $\mathcal{N}_t^{\rm n = 12}$ is the estimated number of timesteps a ($n=12$)-only DBH equiilbration run would take to achieve the same number of swaps/monomer that our method achieves in $\mathcal{N}_t$ timesteps.  $R^*$ is the estimated speedup over the protocol discussed in Ref.\ \cite{auhl03} wherein the Lennard-Jones interactions are force-capped and equal times are devoted to stages with $k\sigma^2/\epsilon = 10,\ 20,\ \rm{and}\ 30$.} 
\begin{tabular}{cccc}
$\kappa/\epsilon$ & Swaps/monomer & $R$ & $R^*$ \\
 0 & 10.04 & 23.0 & 2.1\\
 0.5 & 11.54 & 28.2&2.6\\
 1 & 10.24 & 30.1 & 2.8 \\
 1.5 & 7.84 & 31.0 &3.0\\
 2 & 5.46 & 31.3 & 3.1\\
 2.5 & 3.62 & 31.5 &3.2\\
 3 & 3.30 & 31.6 & 3.1 \\
 3.5 & 3.16 & 31.8&3.2 \\
 4 & 2.94 & 33.1 & 3.2\\
 4.5 & 2.84 & 34.8 &3.2\\
 5 & 2.76 & 37.4 & 3.1\\
 5.5 & 2.72 & 40.1 & 3.0\\
\end{tabular}
\end{ruledtabular}
\label{tab:swapspeed}
\end{table}

Previous implementations of DBH\cite{auhl03,svaneborg16} have employed force-capped Lennard-Jones potentials that allow deeper chain interpenetration and reduce the energy barrier for chain crossing.
The original implementation \cite{auhl03} also employed a ramping procedure for $k_{\rm FENE}$, stepping it from $10\epsilon\sigma^{-2}$ to $30\epsilon\sigma^{-2}$ over the course of equilibration runs.
Both procedures reduce the estimated speedups $R$ achieved by our method, by an amount that depends on implementation details.
Table \ref{tab:swapspeed} also presents the estimated speedups $R^*$ over an implementation wherein Lennard-Jones forces are capped at their values for 
$r = .8\sigma$ and equal times are spent on the abovementioned three $k_{\rm FENE}$ values.
These estimates are smaller, typically $\sim 3$ for semiflexible chains.
However, using lower $k_{\rm FENE}$ without also reducing $n$ substantially increases $\ell_0$, impeding equilibration on all length scales, e.g.\ by producing massive chain retraction when $k_{\rm FENE}$ is increased.
Thus we assert that these values of $R^*$ are in fact loose lower bounds for the actual speedups achieved by our algorithm.
Moreover, since well-entangled semiflexible chains require a substantial amount of time to equilibrate, even a speedup factor of 3 is valuable.
For example, equilibrating our $N_{\rm ch} = 1000$ $N = 400$ melts for an additional 200 million timesteps [i.e., $\mathcal{N}_t (R^* - 1)$ for $\kappa = 5.5\epsilon$] would require hundreds of hours and thousands of CPU-core hours on a typical cluster node.

Finally, note that a separate set of simulations that followed the same protocol but were half as long (i.e., have $f = f_0/2 = 23.1677$ in Eq.\ \ref{eq:taufix}) produced $N_{\rm eK}$ values that were systematically lower than those reported above, but by only about 1 percent.
From this data, we can conclude that the minimum number of swaps per monomer required to fully equilibrate well-entangled semiflexible bead-spring melts is about 2.5.

\section{Equilibrium structure of semiflexible Kremer-Grest melts}
\label{sec:equilstruct}

\begin{figure}[h]
\includegraphics[width=3in]{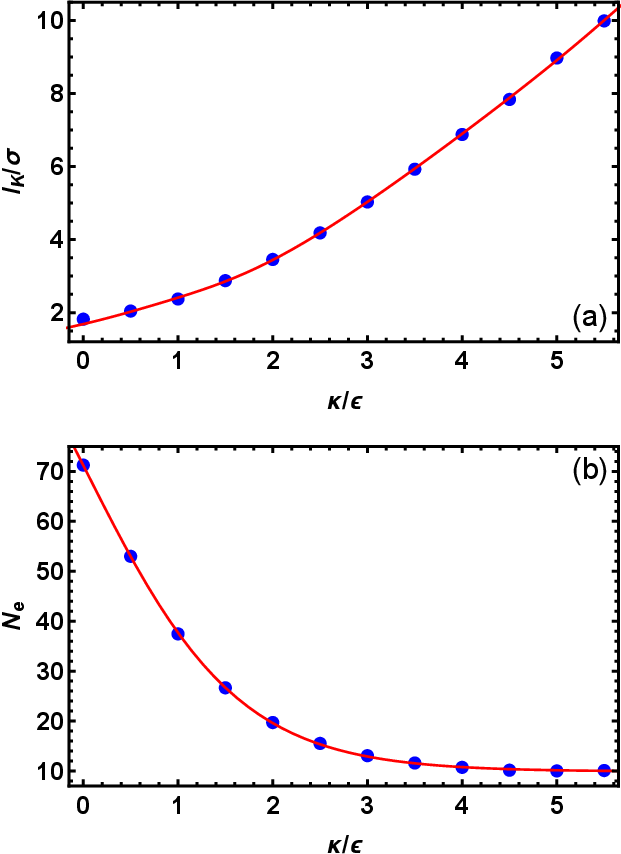}
\caption{Kuhn length and  estimated entanglement length of semiflexible $N = 400$ Kremer-Grest melts. Blue symbols show MD/PPA data obtained using Eq.\ \ref{eq:carNe} and the measured values of $\langle \ell_K(\kappa) \rangle$ while the red curves  in panels (a-b) respectively show Eqs.\ \ref{eq:ourkuhn}-\ref{eq:noeofkap}.}
\label{fig:NeK}
\end{figure}

Figure \ref{fig:NeK} summarizes the structure and entanglement of our fully equilibrated  $N = 400$ melts.
 All results are averaged over the 21 $\tau_e$-separated snapshots from the extended $n = 12$ runs for each of the 10 independently prepared samples, i.e.\ over 210 statistically independent snapshots for each $\kappa$.
Panel (a) shows how $\ell_K$ increases with $\kappa$.
For the $N = 400$ chains considered here, Eq.\ \ref{eq:svanekuhn}  underpredicts $\ell_K$ by $\sim 3\%$ for $\kappa = 0$, is accurate to within $\sim \pm 1\%$ for  $\epsilon < \kappa < 5\epsilon$, and underpredicts $\ell_K$ by $\sim 3\%$  for $\kappa \geq 5\epsilon$.
We find that $\ell_K$ is better fit by the very similar equation
\begin{equation}
\begin{array}{lcl}
\displaystyle\frac{\ell_K}{\ell_0} & = & \displaystyle\frac{2 \kappa/\epsilon + \exp(-2 \kappa/\epsilon)  -1 }{1 - \exp(-2 \kappa/\epsilon) (2 \kappa/\epsilon + 1)} \\
\\
& & +\ 0.364\left[ \tanh( 0.241\kappa^2/\epsilon^2 - 1.73\kappa/\epsilon  + 2.08) + 1 \right].
\end{array}
\label{eq:ourkuhn}
\end{equation}
This expression underpredicts $\ell_K$ by $\sim 3\%$ for $\kappa = 0$,\cite{eq12} but it is accurate to within $< 1.5\%$ for all $\kappa > 0$.

Panel (b) shows results for $N_e(\kappa) = N_{\rm eK}(\kappa)\ell_K(\kappa)/\ell_0$.
We find that $N_e(\kappa)$ is well fit by
\begin{equation}
N_e(\kappa) = 71.2 - 61.3\tanh\left( \displaystyle\frac{\kappa}{1.631\epsilon} \right).
\label{eq:noeofkap}
\end{equation}
This formula for $N_e(\kappa)$ has not been purged of the finite-$N$ errors that are present in many $N_e$-estimators.\cite{hoy09}
Specifically, Eq.\ \ref{eq:carNe} includes a systematic $\mathcal{O}(N^{-1})$ error arising from improper treatment of chain ends that causes it to predict $N_{\rm eK} = 0$ for unentangled chains with $L_{pp} = L$.
To correct for this and extrapolate to the $N \to \infty$ limit as is desired when using $N_e$-estimators,\cite{hoy09} we prepared equilibrated melts with $N = 100, 133, 200$ for all $\kappa$ by cutting the parent $N = 400$ chains into pieces\cite{footN133} and then continuing $n = 12$ equilibration for an additional $2 \tau_e$.
We also prepared equilibrated $N = 533,800$ melts for $\kappa \leq \epsilon$ using the same procedure described in Section \ref{sec:methods}.
Then we performed extended $n = 12$ runs of duration $20\tau_e$ and analyzed the entanglement of $\tau_e$-separated snapshots as described above.

\begin{figure}[h]
\includegraphics[width=3in]{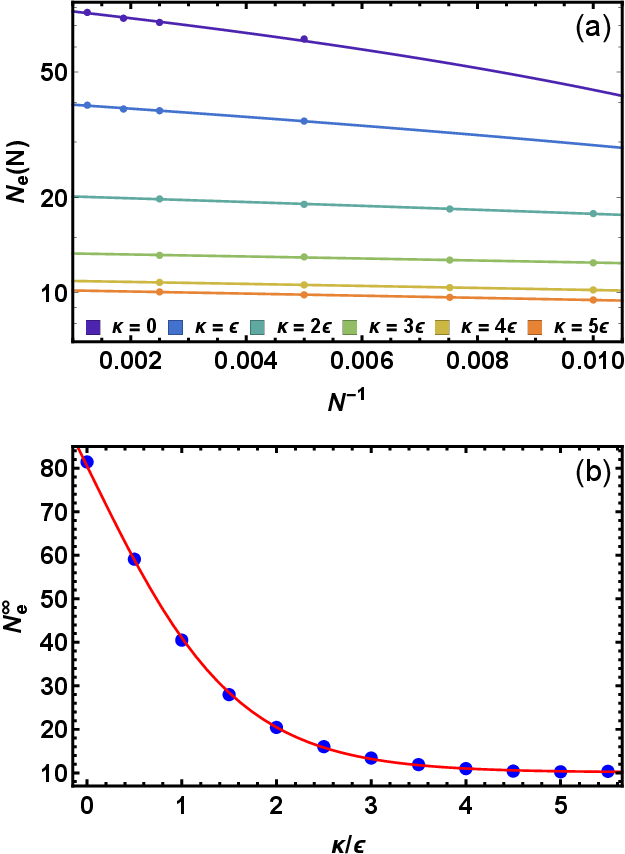}
\caption{Extrapolated $N \to \infty$ entanglement lengths of Kremer-Grest melts.  Panel (a) illustrates the estimated $N_e(N)$ (from Eq,\ \ref{eq:carNe}) and fits to Eq.\ \ref{eq:ONcorrection} for selected $\kappa$, while panel (b) shows the $N_e^\infty(\kappa)$ and Eq.\ \ref{eq:neinfofkap}.
Data are averaged over 10 independently prepared systems for $N \geq 400$, 5 for $N = 200$, 4 for $N = 133$, and 3 for $N = 100$.  For $\kappa \leq \epsilon$, data for $N < 200$ are not included in the fits to Eq.\ \ref{eq:ONcorrection} because they exhibit non-leading-order corrections to $N_e^{est}(\kappa,N)$.}
\label{fig:NeInf}
\end{figure}

Figure \ref{fig:NeInf} summarizes our findings.
Panel (a) shows the $N_e$ values estimated from Eq,\ \ref{eq:carNe} as well as fits to 
\begin{equation}
N_e^{est}(\kappa,N) = N_e^\infty(\kappa) - \displaystyle\frac{b(\kappa)}{N}
\label{eq:ONcorrection}
\end{equation}
for selected $\kappa$.
This empirical form accurately describes the data for all $\kappa$  and $N \gtrsim 2 N_e^\infty(\kappa)$ considered here, and should remain accurate in the $N\to\infty$ limit.\cite{hoy09}
Panel (b) shows results for $N_e^\infty(\kappa)$.
We find that $N_e^\infty(\kappa)$ is well fit by
\begin{equation}
N_e^\infty(\kappa) = 80.5 - 70.4\tanh\left( \displaystyle\frac{\kappa}{1.579\epsilon} \right).
\label{eq:neinfofkap}
\end{equation}
Note that other topological-analysis methods such as Z, CReTA, and thin-chain PPA \cite{kroger05,tzoumanekas06,hoy07b} will, in general, give different $L_{\rm pp}(\kappa)$ and thus quantitatively different formulae for $N_e(\kappa)$.

The plateau in $N_e^\infty$ for $4\epsilon \lesssim \kappa \lesssim 5.5\epsilon$ is consistent with both our earlier results for $\rho = .7\sigma^{-3}$ systems\cite{hoy20}
and Bobbili and Milner's results for entangled Olympic-ring polymer melts.\cite{bobbili20}
More generally, it is consistent with Milner's suggestion\cite{milner20} that entanglement is maximal and has a plateau in the semiflexible-chain regime where entangled segments approximately correspond to Kuhn segments ($N_{\rm eK} \simeq 1$).

\section{Discussion and Conclusions}

Bead-spring polymer melts remain of substantial interest owing both to their utility for elucidating general features of polymer rheology that remain poorly understood\cite{oconnor19,oconnor20} as well as their ability to map quantitatively to common synthetic polymers.\cite{svaneborg20,everaers20}
Semiflexible bead-spring melts have attracted renewed interest owing to their ability to probe the poorly understood crossover regime between flexible and stiff entanglement\cite{hoy20,milner20,bobbili20} and to their potential applicability for modeling recently developed conjugated polymers which lie in this regime and are under intensive study for their potential use in flexible electronic circuits.\cite{xie20,fenton21}

Here we developed and described a method that is suitable for preparing equilibrated well-entangled ($N \simeq 40 N_e$) semiflexible bead-spring polymer melts for chain stiffnesses up to the isotropic-nematic transition.
Our method combinines two previously employed methods (core-softening\cite{sliozberg12,bobbili20} and double bridging\cite{karayiannis02,karayiannis02b,auhl03}) in a novel, controlled fashion.
It provides a speedup by a factor of at least 3 over standard DBH\cite{auhl03} and can be straightforwardly improved upon in several different ways. 
For example, use of Monte Carlo prepacking schemes\cite{generationMK,auhl03,moreira15,svaneborg16,hsu20} that reduce local density fluctuations in the initial states would bring these states closer to equilibrium and reduce $\tau_{\rm eq}$.
Combining such schemes with a protocol that uses a more sophisticated criterion for optimizing the initial states' chain statistics\cite{generationMK,moreira15,generationMKupdate,hsu20} should prove fruitful.
Furthermore, since we did not attempt to optimize our choices of $\Xi$ (Eq.\ \ref{eq:nset}) and $\tau_i(\kappa)$ (Eqs.\ \ref{eq:scheme1portions}-\ref{eq:scheme2portions}), further refinement might provide a significant additional speedup.
Note that our method is suitable for generating coarse-grained configurations which can be used in conjunction with configurational-backmapping methods\cite{zhang14,zhang15,zhang19} to generate equilibrated well-entangled semiflexible atomistic or united-atom-model  polymer melts.

We conclude with a remark on the relation of the equilibration method described herein to methods that employ core-softening but not topology-changing.\cite{sliozberg12,bobbili20}   
The latter \textit{are} capable of equilibrating semiflexible polymer melts using the computational resources available to typical academic research groups.
They are likely the best currently available method that is straightforwardly applicable to branched or ring polymer melts where DBH-like Monte Carlo moves\cite{bisbee11,qin11,qin14,qin16} are chain-architecture-specific and challenging to implement.
They are also readily applicable to atomistic models with stiff bond-angular interactions.
On the other hand, for linear bead-spring polymers, they produce substantially lower chain mobilities and correspondingly slower convergence of $N_e$ than are achieved when topology-changing moves are added.
Further details are given in Appendix B.

The code used in this work is being made available as part of LAMMPS' EXTRA-PAIR and EXTRA-MOLECULE packages (https://lammps.sandia.gov/), and LAMMPS scripts for our method as well as the equilibrated melts described above are available on our website (http://labs.cas.usf.edu/softmattertheory/).

\section{Acknowledgements}

This material is based upon work supported by the National Science Foundation under Grant No.\ DMR-1555242.
We thank Martin Kr{\"o}ger and Carsten Svaneborg for numerous helpful discussions.

The authors have no conflicts to disclose.

\begin{appendix}

\section{Verification of multiscale structural equilibration}

\begin{figure}[h]
\includegraphics[width=3in]{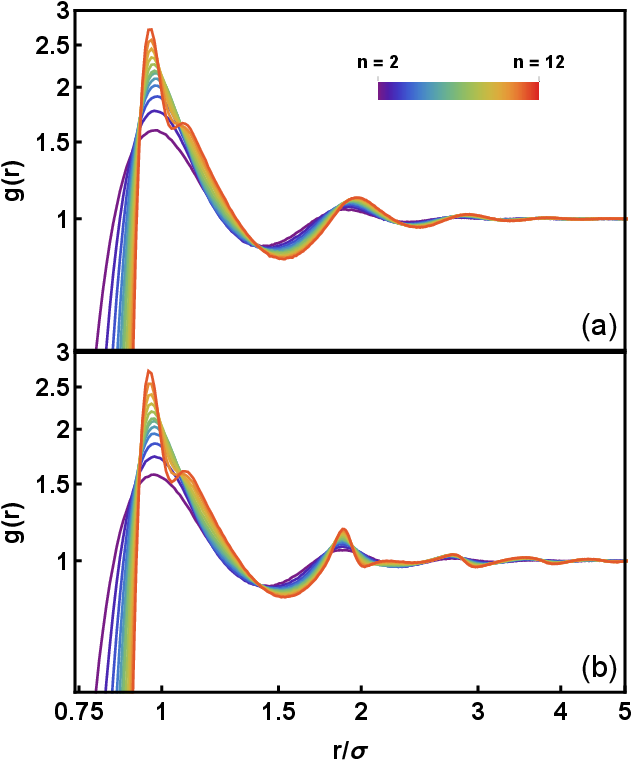}
\caption{Evolution of the pair correlation function $g(r)$ for melts with $\kappa=0.0$ [panel (a)] and $\kappa=5.5\epsilon$  [panel (b)] during equilibration. The color-bar legend represents the exponent $n$ in the $U_{\rm MLJ}^n$ potential (Eq.\ 6), and data is shown for the configurations at the end of each $n$-step (Eqs.\ 11, 19-20).}  
\label{fig:gofr}
\end{figure}

In this Appendix, we demonstrate that the $N = 400$ melts prepared using the procedure described above are well-equilibrated on all length scales.
We show that the pair correlation functions [$g(r)$] and static structure factors [$S(q)$] of our most flexible ($\kappa=0$) and stiffest ($\kappa=5.5\epsilon$) melts evolve monotonically with increasing $n$ and have converged by the end of the $n = 12$ stage of their equilibration runs. 
Although $g(r)$ and $S(q)$  are each other's Fourier transforms and hence formally contain the same information, visual inspection of both is useful because each highlights aspects of the progress towards equilibrium that are not apparent from inspection of the other.
All data is averaged over the 10 independently prepared samples described in Section \ref{sec:perf}.

Figure \ref{fig:gofr} illustrates the evolution of $g(r)$:\ colors indicate data from the end of each $n$-step.
As $n$ increases and $U_{\rm MLJ}^n(r)$ and $U_{\rm bond}^n(r)$ approach their final functional forms, the peaks in $g(r)$ gradually sharpen.
The broad initial peak for low $n$ splits in two:\ the larger-$n$ peaks at at $r \simeq \ell_0 \simeq 0.97\sigma$ and $r \simeq 2^{1/6}\sigma$ respectively correspond to covalently bonded neighbors and nearest nonbonded neighbors.
This splitting roughly corresponds to cessation of chain crossing; see Fig.\ 2.
At larger $n$, higher-$\kappa$ systems have sharper peaks at larger $r$ owing to their larger $\ell_K$.
We emphasize that equilibration  algorithms that employ only core-softening (i.e.\ do not employ topology-switching) would require substantially longer runtimes to achieve  the same convergence to equilibrium over the range $9 \lesssim n \leq 12$, because in the absence of topology-switching chains would be constrained to their rheological tubes and hence reptate rather than diffusing more freely.

\begin{figure}
\includegraphics[width=3in]{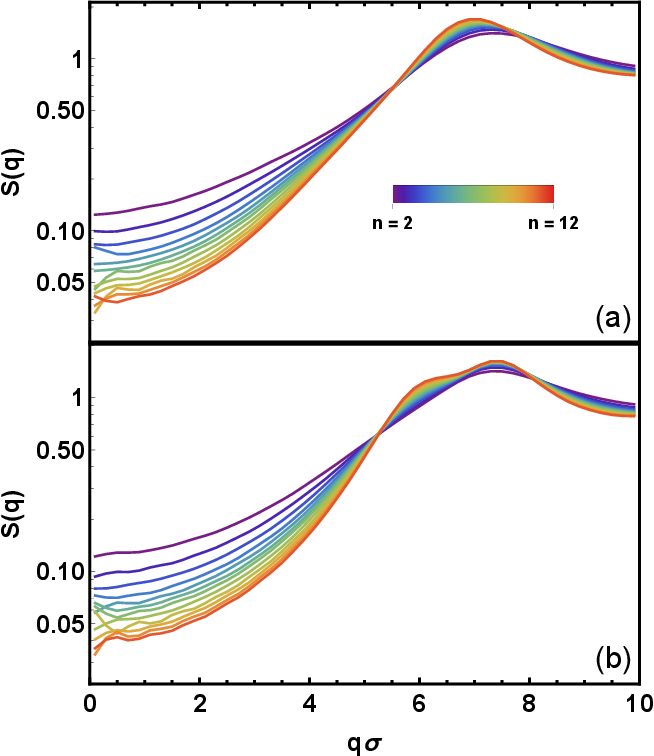}
\caption{Evolution of the static structure factor $S(q)$ for $\kappa=0.0$ [panel (a)] and $\kappa =5.5 \epsilon$ [panel (b)] during equilibration.  Colors are the same as in Fig.\ \ref{fig:gofr}.}
\label{fig:soq}
\end{figure}

Figure \ref{fig:soq} illustrates the evolution of $S(q)$.
$\lim_{q \to 0} S(q)$ decreases rapidly with increasing $n$ as long-wavelength density fluctuations anneal out; this decrease is a key indicator of equilibration \cite{generationMK,auhl03,zhang14,moreira15,svaneborg16,lemarchand19}. 
On the other hand, for $5.0 \sigma^{-1} < q < 8.0 \sigma^{-1}$, $S(q)$ increases with increasing n. 
This trend is aligned with the sharpening of the first peak in $g(r)$ at $r =\ell_0$.
Finally, for $q>8.0\sigma^{-1}$, $S(q)$ decreases with increasing $n$ as intermonomer distances substantially below $\ell_0$ become increasingly unlikely.
Again we emphasize that this progression towards equilibrium structure (especially at low $q$) must necessarily be slower in the absence of topology switching.

At the end of the equilibration runs, the primary peaks in $S(q)$ are respectively located at $q = 6.9 \sigma^{-1}$ and $q = 7.4 \sigma^{-1}$ for $\kappa = 0.0$ and $\kappa = 5.5 \epsilon$.
Intriguingly, the $\kappa=5.5\epsilon$ distribution exhibits a prominent secondary peak at $q=5.6\sigma^{-1}$ that is not present for flexible chains.
Semiflexible-chain monomers are far more likely to be surrounded by interchain (as opposed to intrachain) neighbors than their flexible-chain counterparts; we believe that this secondary peak is a signature of the concomitant increased local interchain ordering.

\section{Comparison to a core-softening-only approach }

As shown in Figs.\ \ref{fig:gofr}-\ref{fig:soq}, the core-softened interactions employed here lead to two-body structure that differs from equilibrium structure over a wide-range of length scales.
Analogous simulations that employed \textit{only} core-softening, i.e.\ simulations that followed the same $2 \leq n \leq 12$ progression described in Section \ref{sec:methods} but did not employ any double-bridging moves, produce a very similar evolution of structure with increasing $n$.
However, as we will show here, chain mobility in these simulations is much lower and the convergence of $N_e$ towards equilibrium is correspondingly slower.

 \begin{figure}
\centering
\includegraphics[width=3in]{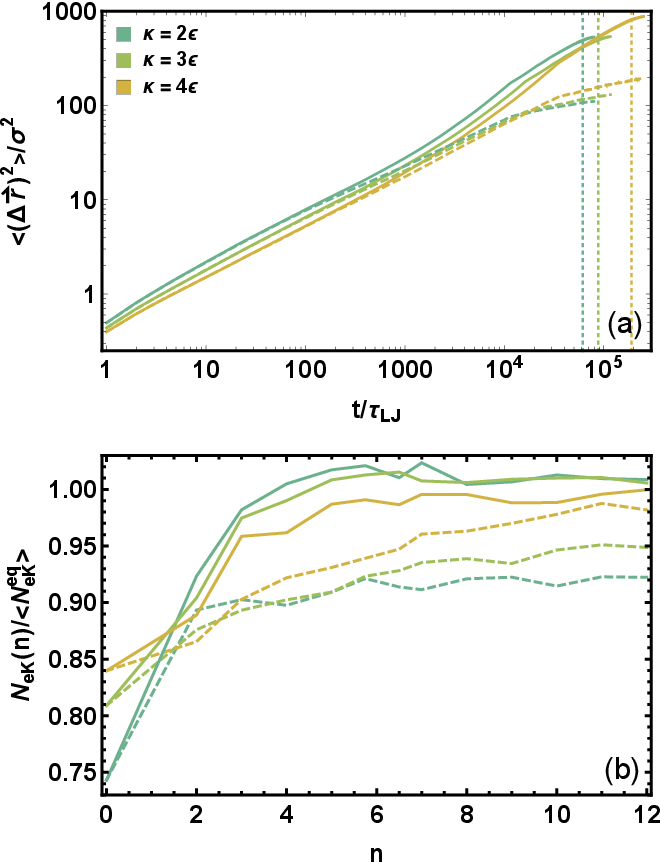}
\caption{Mobility and convergence of $N_e$ to equilibrium for $N = 400$, with (solid curves) and without (dashed curves) double-bridging moves.  Panel (a) shows monomeric mean-squared displacements starting from the beginning of the $n = 2$ step.  Vertical dotted lines indicate the beginning of the $n = 9$ step and roughly mark the cessation of chain crossing.   Panel (b) shows  $N_{\rm eK}(n)/\langle N_{\rm eK}^{\rm eq} \rangle$, where $\langle N_{\rm eK}^{\rm eq} \rangle$, where are the ensemble-averaged equilibrium values; the final values for the combined protocol differ slightly from unity because these results are not ensemble-averaged.  Results for $\kappa < 2\epsilon$ and $\kappa > 4\epsilon$ are consistent with the trends shown here and are omitted for clarity.}
\label{fig:appendB}
\end{figure}

Figure \ref{fig:appendB} illustrates how the absence of double-bridging moves can greatly slow equilibration even when chain crossing is allowed.
Panel (a) compares the core-softening-only vs.\ combined-approach monomeric mean-squared displacements in semiflexible melts with $2\epsilon \leq \kappa \leq 4\epsilon$.
Here the reference ($t = 0$) state is the beginning of the $n = 2$ step (Section \ref{sec:methods}).
For all chain stiffnesses, the mobilities are comparable until $\langle (\Delta \vec{r})^2 \rangle$ exceeds $\sim 10\sigma^2$.
Subsequently, chains are far more mobile when double bridging is employed, and $\langle (\Delta \vec{r})^2 \rangle$ values in these runs are roughly an order magnitude larger than their core-softening-only counterparts by the end of the $n = 12$ stage.
For intermediate $t$ and $n$, the effective power law $\gamma(t) = d \ln[ \langle (\Delta \vec{r})^2 \rangle]/dt$ of the subdiffusion in the core-softening-only simulations is substantially lower.
We emphasize that this lower mobility occurs even though chains are crossing (presumably at approximately the same rate since the barriers to crossing depend only on $n$; see Fig.\ \ref{fig:pots}) in both sets of simulations.
It indicates that chains in the core-softening-only simulations remain at least somewhat confined to their tubes.

Panel (b) shows the ratio of $N_{\rm eK}$ at the each $n$-step to the equilibrium $\langle N_{\rm eK} \rangle$.
As in Fig.\ \ref{fig:convwithn}(b), $N_{\rm eK}/\langle N_{\rm eK}^{\rm eq} \rangle$ converges to unity (within our statistical uncertainties of $\sim 0.5\%$) by the end of the combined-protocol runs.
For the core-softening only runs, however, the final $N_{\rm eK}/\langle N_{\rm eK}^{\rm eq}  \rangle$ values range from $0.92$ to $0.98$.
In general, parameters such as $N_e$ that depend on higher-order structural correlations on some length scale $s$ cannot be expected to have equilibrated until typical monomers have diffused by more than $s$.\cite{moreira15,svaneborg16}
For $N_e$, one expects $s \geq a$, where the tube diameter $a \simeq \sqrt{N_e \ell_0 \ell_K}$.
Here $7\sigma < a < 9\sigma$ for all $\kappa \geq \epsilon$; thus the poorer convergence of $N_e$  in the core-softening-only runs is unsurprising.
One might argue that this slower convergence is an artifact of our gradual stepping of the interactions from $n = 2$ to $n = 12$ (Eqs.\ \ref{eq:scheme1portions}-\ref{eq:scheme2portions})  and that spending larger portions of the runs at small $n$ would both eliminate the differences shown in Fig.\ \ref{fig:appendB} and allow the simpler core-softening-only approach to equilibrate systems in the same amount of CPU time.
However, as shown in panel (a), the mobilities diverged well \textit{before} the $n = 2$ portions of the runs end at $1.15\cdot 10^4\tau \leq \tau_i(\kappa) \leq 3.56\cdot 10^4 \tau$. 
This suggests that even $n = 2$-only runs of nearly the same length followed by short $n = 12$ runs would not have equilibrated $N_e$ within the same amount of time.

\end{appendix}


\begin{thebibliography}{53}
\bibitem [{mec()}]{mechvstherm}%
  \BibitemOpen
  \href@noop {} {}\bibinfo {note} {Here $\ell_0$ is the bond length for which
  $\partial U_{\rm bond}/\partial r = 0$. Thermodynamic-equilibrium bond
  lengths are slightly larger, and depend on $\rho$ and $T$.}\BibitemShut
  {Stop}%
\bibitem [{eq1()}]{eq12}%
  \BibitemOpen
  \href@noop {} {}\bibinfo {note} {Equations \ref{eq:svanekuhn} and
  \ref{eq:ourkuhn} are well-defined only for $\kappa > 0$; for $\kappa = 0$
  their first terms must be replaced by $1$.\cite{svaneborg20}}\BibitemShut
  {NoStop}%
\bibitem [{foo()}]{footN133}%
  \BibitemOpen
  \href@noop {} {}\bibinfo {note} {The ``N = 133'' systems actually have
  $\langle N \rangle = 400/3$; they were produced by cutting each $N = 400$
  parent chain into two $N = 133$ chain and one $N = 134$ chain.}\BibitemShut
  {Stop}%
\end{thebibliography}

%

\end{document}